\documentclass[a4paper,11pt,headings=big,DIV=12]{scrartcl}
\pdfoutput=1

\usepackage{array} % for centering in p{} columns
\usepackage{booktabs} % for improved table lines
\usepackage{caption}

\usepackage[a4paper, top=1.2in, bottom=1.2in, left=1.1in, right=1.1in]{geometry}
\usepackage[utf8]{inputenc}
\usepackage{amsmath, amssymb}
\usepackage{color}
\usepackage[dvipsnames, table]{xcolor}
\definecolor{blue}{rgb}{0,0,0.5}
\usepackage[colorlinks,linkcolor=blue,urlcolor=blue,citecolor=blue]{hyperref}
\usepackage{graphicx}
\graphicspath{{./figs/}}
\usepackage{slashed}
\usepackage{cite}
\usepackage{enumitem}
\usepackage{bm}
\usepackage{physics}
\usepackage{xspace}
\usepackage[T1]{fontenc}
\usepackage{lmodern}
\usepackage{multirow}
\usepackage{makecell}
\usepackage{booktabs, bm}

\newcommand{\be}{\begin{equation}}
\newcommand{\ee}{\end{equation}}
\newcommand{\bea}{\begin{eqnarray}}
\newcommand{\eea}{\end{eqnarray}}

\newcommand{\mc}{\mathcal}

\newcommand{\noi}{\noindent}

\newcommand{\MeV}{{\rm MeV}}
\newcommand{\GeV}{{\rm GeV}}
\newcommand{\kpc}{{\rm kpc}}
\newcommand{\sat}{{\rm sat}}
\newcommand{\cm}{{\rm cm}}
\newcommand{\aqua}{\rm H_2 O}
\def\refeq#1{Eq.~(\ref{#1})}

\def\reftab#1{Table~\ref{#1}}
\def\reftabs#1#2{Tables~\ref{#1}-\ref{#2}}
\def\reffig#1{Fig.~\ref{#1}}
\def\reffigs#1#2{Figs.~\ref{#1}-\ref{#2}}
\def\refapp#1{Appendix~\ref{#1}}
\def\refsec#1{Sec.~\ref{#1}}

\def\de{\partial}

\def\hc{{\rm h.c.}}

\def\dof{{\rm dof}}

\def\DDmod{{\tt DD2}\xspace}
\def\SFmod{{\tt SFHo}\xspace}

\def\close{\sim}

\newcommand{\Bimba}{B_i\,M\to B_f\,a}
\newcommand{\Biba}{B_i\,\to B_f\,a}

\newcommand{\Cann}{C_{ann}}
\newcommand{\Capp}{C_{app}}

\newcommand{\df}{\mathcal{F}} % symbol for distribution functions
\DeclareOldFontCommand{\rm}{\normalfont\rmfamily}{\mathrm}
\DeclareOldFontCommand{\sf}{\normalfont\sffamily}{\mathsf}
\DeclareOldFontCommand{\tt}{\normalfont\ttfamily}{\mathtt}
\DeclareOldFontCommand{\bf}{\normalfont\bfseries}{\mathbf}
\DeclareOldFontCommand{\it}{\normalfont\itshape}{\mathit}
\DeclareOldFontCommand{\sl}{\normalfont\slshape}{\@nomath\sl}
\DeclareOldFontCommand{\sc}{\normalfont\scshape}{\@nomath\sc}

\newcommand{\che}{\v{C}herenkov\xspace}
\newcommand{\cgg}{c_{GG}}

\def\chep{X} % symbol for the Cherenkov parton
\def\axnu{\alpha} % symbol for, interchangeably, the axion or the neutrino
\newcommand{\laR}{\lambdaR}
\def\lambdaR{\lambda_{\pi^0}}

\newcommand{\scherer}{Scherer:2002tk}
\newcommand{\georgi}{Georgi:1986df}
\newcommand{\neubertPRL}{Bauer:2021wjo}

\newcommand{\mael}{Cavan-Piton:2024ayu}
\newcommand{\lucente}{Lucente:2022vuo}
\newcommand{\caravi}{Caputo:2021rux}
\newcommand{\cajaravi}{Caputo:2022mah}
\newcommand{\pdg}{ParticleDataGroup:2024cfk}
\newcommand{\chacho}{Chang:1993gm}
\newcommand{\diluzio}{DiLuzio:2020wdo}
\newcommand{\vonk}{Vonk:2021sit}
\newcommand{\ddref}{Hempel:2009mc}
\newcommand{\sfref}{Steiner:2012rk}
\newcommand{\carenza}{Carenza:2020cis}

\begin{document}

%%%% TITLE PAGE
\begin{flushright}
\small
LAPTH-011/25
\end{flushright}
%\vskip0.5cm

\begin{center}
%%%% TITLE
{\sffamily \bfseries \LARGE \boldmath
Probing the general axion-nucleon interaction\\[0.3cm]
in water \v{C}herenkov experiments}\\[0.8 cm]
%%%% AUTHORS
{\normalsize \sffamily \bfseries Maël Cavan-Piton$^1$, Diego Guadagnoli$^1$,Axel Iohner$^1$,\\[0.1cm]Pablo Fernández-Menéndez$^2$, Ludovico Vittorio$^{1,3}$} \\[0.5 cm]
\small
$^1${\em LAPTh, Universit\'{e} Savoie Mont-Blanc et CNRS, Annecy, France}
\\
[0.1cm]
$^2${\em Donostia International Physics Center DIPC, San Sebasti\'an/Donostia, E-20018, Spain}
\\
[0.1cm]
$^3${\em Sapienza Università di Roma and INFN, Sezione di Roma, Italy}
\end{center}

\medskip

\begin{abstract}
\noi We consider an axion flux on Earth consistent with emission from the Supernova explosion SN~1987A. Using Chiral Perturbation Theory augmented with an axion, we calculate the energy spectrum of $a + N \to N + \gamma$ as well as $a + N \to N + \pi^0$, where $N$ denotes a nucleon in a water tank, such as the one planned for the Hyper-Kamiokande neutrino detection facility.
Our calculations assume the most general axion-quark interactions, with couplings constrained either solely by experimental data, or by specific theory scenarios.

We find that even for the QCD axion---whose interaction strength with matter is at its weakest as compared with axion-like particles---the expected \che-light spectrum from neutrino-nucleon interactions is modified in a potentially detectable way. Furthermore, detectability appears significantly more promising for the $N + \pi^0$ final state, as its spectrum peaks an order of magnitude higher and at energies twice as large compared to the $N + \gamma$ counterpart. Given the rarity of SN events where both the neutrino and the hypothetical axion burst are detectable, we emphasize the importance of identifying additional mechanisms that could enhance such signals.
\end{abstract}

\newpage

\tableofcontents

\section{Introduction}

Axions are hypothetical but highly plausible particles, as they offer an elegant solution to key shortcomings of the Standard Model. These include both theoretical issues---such as the surprisingly tiny coupling observed for the dimension-4 operator that violates the $CP$ symmetry in strong interactions, with no apparent reason for this suppression, not even anthropic \cite{Ubaldi:2008nf,Dine:2018glh}---and phenomenological challenges, in particular an explanation of the matter fraction attributed to Dark Matter \cite{Preskill:1982cy,Abbott:1982af,Dine:1982ah}.

In Supernova (SN) explosions, the cooling mechanism leading to the formation of a young neutron star is known to provide one of the most powerful existing constraints on exotic sources of cooling like axions, on top of the known emission of neutrinos. Once produced by the SN, the axion flux may accordingly be detected on Earth through facilities designed to detect neutrinos, for example the foreseen Hyper-Kamiokande water \che detector (HK)~\cite{Abe:2011ts, Hyper-Kamiokande:2018ofw}. The main challenge is to disentangle the two fluxes.

Several papers in the literature have explored different physics mechanisms for axion detection at neutrino \che experiments. Examples we are aware of include: {\em (a)} The coupling of the QCD axion to matter induced by the nucleon electric dipole moment (EDM) portal interaction, a model-independent feature of QCD axions, as discussed in \cite{\lucente}; {\em (b)} Investigations into the axion-photon-photon coupling and the nucleon EDM portal interaction for axion-like particles (ALPs) produced from primordial black hole evaporation \cite{Li:2022xqh}; {\em (c)} Studies of the coupling responsible for axion photo-production through the anomaly-induced Wess-Zumino-Witten term, see \cite{Chakraborty:2024tyx}; {\em (d)} Constraints on the axion-photon and the axion-electron couplings for ALPs originating from neutrino-boosted fluxes \cite{Chao:2023dwc}; {\em (e)} Studies of the couplings of axions with electrons, protons, and neutrons for both solar and supernova (SN) axions, focusing on processes such as inverse Compton scattering, the axio-electric effect, and axion-induced pair production \cite{Arias-Aragon:2024gdz}; {\em (f)} The coupling of glueball ALPs with protons, as investigated in \cite{Carenza:2024avj}; {\em (g)} The coupling between protons and ALPs from a diffuse galactic flux via photo-production, as discussed in \cite{Alonso-Gonzalez:2024ems, Alonso-Gonzalez:2024spi}; {\em (h)} axion-absorption induced excitation of oxygen nuclei, leading to the emission of de-excitation photons, potentially providing detectable signals at \che detectors \cite{Engel:1990zd, Carenza:2023wsm}. All these mechanisms underscore the versatility of \che experiments in probing a wide range of axion interactions, as well as their astrophysical implications.

In this paper, we consider the most general interactions of the QCD axion with nucleons. Using the Noether procedure, these interactions can be defined as an extension of Chiral Perturbation Theory for nucleons, with their strengths inheriting from the couplings of the axion to quarks \cite{\georgi}. These couplings are typically referred to as model-dependent, to distinguish them from the so-called ``anomalous'' couplings of the axion, whose presence is unavoidable. Axion-quark couplings are not fundamentally required to be absent, and, like Yukawa couplings, will in general be present. In fact, they (including the allowed flavor-violating ones) may be large within the limits of existing constraints and, again like Yukawa couplings, may lead to detectable signals. Using the thus-defined nucleonic chiral Lagrangian coupled to axions, we discuss axion-absorption processes in water that produce final-state ``\che partons'', i.e. initiators of \che-light cascades. We find that $\pi^0$ particles are especially promising, because their number spectrum has a higher peak and is more clearly separated in energy than the photon counterpart. In both cases, peak heights and energy separations are compared to the reference \che spectrum expected from neutrino-absorption events.

The structure of this work is as follows. In \refsec{sec:theory_setup}, we present our theoretical framework, specifically the nucleonic chiral Lagrangian augmented with an axion, and outline different approaches to constraining its couplings. In \refsec{sec:detection}, we describe the calculation of number spectra for processes in which an axion is absorbed and produces \che light, detectable at facilities such as HK. Useful analytical expressions are provided in \refapp{app:sigma}. We then use these number spectra to derive experimental bounds on the couplings from astrophysical observations, which are collected in \refapp{app:bounds}.
In \refsec{sec:spectra}, we compare axion-absorption-induced spectra with their neutrino counterparts, both at the physical level and after applying detector-response corrections, the latter discussion being supplemented by tables in \refapp{app:T_30_40}.
We analyze the dependence of the expected event rate on axion-matter couplings and explore the connection between this rate and the number of potential SN candidates.
Finally, \refsec{sec:outlook} provides an outlook.

\section{Theory setup} \label{sec:theory_setup}

Given that the axion is well below the hadronization scale, its interactions with light quarks have been consistently formulated in the context of Chiral Perturbation Theory \cite{\georgi}. Without lack of generality, the underlying interaction may be written as \cite{\georgi,\neubertPRL}
\be
\label{eq:Laqq}
\mc L_{aqq} ~\equiv~ \frac{\de_\mu a}{f_a} 
\left(
\bar q \, \gamma_L^\mu \hat{{k}}_{L}{(a)} \, q + 
\bar q \, \gamma_R^\mu \hat{{k}}_{R}{(a)} \, q
\right)~,
\ee
with $q = (u, d, s)^T$ and $\gamma_{L,R}^\mu = \gamma^\mu(1\mp\gamma_5)/2$. Before matching this interaction with the axion-hadron interaction Lagrangian, the $a G \tilde G$ coupling is traded for an axion-dependent phase in the quark mass matrix $M_q$, via the quark-field redefinition \cite{\georgi} $q \to \exp{-i \cgg \frac{a}{f_a}(\delta_q + \gamma_5 \kappa_q)} q$, with $\delta_q$ and $\kappa_q$ hermitian matrices, and $\cgg$ the axion-gluon coupling in the normalization $\mc L_{a} \supset \cgg \frac{\alpha_s}{4 \pi} \frac{a}{f_a} G_{\mu \nu} \tilde G^{\mu \nu}$, with $\tilde G_a^{\mu \nu} \equiv \frac{1}{2}\epsilon^{\mu \nu \rho \sigma} G_{\rho \sigma, a}$~\cite{\neubertPRL}. This redefinition immediately implies
\be
\label{eq:hatkRL}
\hat k_{R,L}(a) = e^{i a / f_a \cgg (\delta \pm \kappa)} \bigl(k_{R,L} + \cgg (\delta \pm \kappa)\bigl) e^{- i a / f_a \cgg (\delta \pm \kappa)}~.
\ee
The dependence on contributions proportional to the $\delta$ and $\kappa$ parameters must cancel in physical observables \cite{\neubertPRL}. The fundamental axion couplings to light quarks are thus encapsulated in the 3-by-3 matrices ${k}_{L,R}$ in \refeq{eq:hatkRL}, whose only non-trivial entries are 11, 22, 33, and 23 (ten real parameters). Similarly as Yukawa couplings, these entries are only constrained by data on axion emission or absorption from hadronic matter. Axion couplings to hadrons inherit from the interaction \refeq{eq:Laqq}. Following the notation in Ref.~\cite{\mael}, and denoting the octet-meson and octet-baryon fields as $U$ and $B$ respectively, we express the axion-hadron interactions as\footnote{We neglect the contribution from the SM chiral Lagrangian mediating weak processes~\cite{Cronin:1967jq,Kambor:1989tz} and augmented with an axion, as we restrict to first-generation matter. Besides, this contribution is generally suppressed by $G_F F_0^2 \sim 10^{-7}$ relative to the interactions we consider \cite{\mael}.}
\be
\label{eq:LaUB}
\mc L_{a U B} ~=~ \frac{\de_\mu a}{f_a} \sum_{b =1}^8
\left( x^b_L(k_L) ~\mc J_L^{b \mu}(U,B) + x^b_R(k_R) ~\mc J_R^{b \mu}(U,B)
\right)~,
\ee
where the explicit form of the $\mc J$ currents and for the $x$ couplings can be read off from Ref.~\cite{\mael}.
In this work we restrict only to first-generation hadrons, i.e. nucleons and pions, and the corresponding interaction terms are compact enough to be written explicitly.
The participating interactions are $\mc L_{a NN}$, $\mc L_{NN \gamma}$, $\mc L_{N N^\prime \pi}$, $\mc L_{a \pi \pi^\prime}$,
and $\mc L_{a \pi N N^\prime}$.
In the formulae to follow we use $k_{V,A} \equiv k_R \pm k_L$, take $m_s \to \infty$ and redefine $m_u = z \, m_d$ in order to make contact with Ref.~\cite{\chacho}.

We take the axion-nucleon interaction $\mc L_{aNN}$ in the usual normalization
\be
\mc L_{a U B} \supset \mc L_{aNN} = \frac{\de_\mu a}{2 f_a} \sum_{N = p,n}  
C_{aNN} \bar N \gamma^\mu \gamma^5 N~,
\ee
with
\be
C_{aNN} ~\equiv~ C^{(D)}_{aNN}~D + C^{(F)}_{aNN}~F + C^{(S)}_{aNN}~S~,
\ee
and find
\be
\label{eq:CaNN}
\begin{aligned}
C^{(D)}_{ann} &=~~ \frac{2}{3}\cgg \frac{2 - z}{1 + z} + \frac{2 (k_A)_{11} - (k_A)_{22} - (k_A)_{33}}{3} ~, \\
C^{(F)}_{ann} &=~~ - 2\cgg \frac{z}{1 + z} - (k_A)_{22} + (k_A)_{33}~, \\
C^{(D)}_{app} &=~~  - \frac{2}{3}\cgg \frac{1 - 2z}{1 + z} - \frac{(k_A)_{11} - 2 (k_A)_{22} + (k_A)_{33}}{3} ~, \\
C^{(F)}_{app} &=~~  -2 \frac{\cgg}{1 + z} - (k_A)_{11} + (k_A)_{33} ~,\\
C^{(S)}_{app} &=~~C^{(S)}_{ann} =~~  \dfrac{2}{3}\cgg + \frac{(k_A)_{11} +(k_A)_{22} + (k_A)_{33}}{3} ~.
\end{aligned}
\ee
These expressions are in agreement with Ref.~\cite{\chacho} if we set $c_{GG} = 1/2$ and identify our $D, F$ couplings with minus those in that reference. This is also consistent with Ref.~\cite{\scherer}, as discussed in Ref.~\cite{\mael}. We also find
\be
\label{eq:CappCann}
\begin{aligned}
	C_{ann}&=  0.012(28) -0.406(34)~(k_A)_{11} + 0.848(26)~(k_A)_{22} -0.035(17)~(k_A)_{33} ~,\\
	C_{app}&= -0.452(28) + 0.848(34)~(k_A)_{11} - 0.406(24)~(k_A)_{22} -0.035(17)~(k_A)_{33} ~,
\end{aligned}
\ee
which is consistent with Ref.~\cite{\diluzio}. This expression is obtained setting $c_{GG}=-1/2$ (note that the axion-dependent quark redefinition has an overall opposite sign in \cite{\diluzio} than in \cite{\chacho,\neubertPRL,\georgi} with which we comply) as well as the relations
\be
\label{eq:DFS}
\begin{aligned}
		D &=-\Delta_u/2+\Delta_d-\Delta_s/2 = -0.813(43)~,\\
		F &=-\Delta_u/2+\Delta_s/2 = -0.441(26)~,\\
		S &=\Delta_u+\Delta_d+\Delta_s = 0.405(62)~,
\end{aligned} 
\ee
which agree with Ref.~\cite{\vonk}.

The $C_{aNN}$ couplings in \refeq{eq:CaNN} are completely general. For the purposes of the present study, we will specialize them to a few cases of interest, as follows.
\begin{itemize} \label{page:models}
\item KSVZ~\cite{Kim:1979if,Shifman:1979if}: in this model the only contributions to the axion-quark-quark couplings $(k_A)_{11}$, $(k_A)_{22}$ and $(k_A)_{33}$ are of $O(10^{-2})$ from running effects~\cite{Choi:2021ign}, implying
\be
C_{ann}\vert_{ \rm KSVZ}= 0.012(28)~,\qquad C_{app}\vert_{\rm KSVZ}=-0.452(28).
\ee
\item DFSZ~\cite{Dine:1981rt,Zhitnitsky:1980tq}: in this model $(k_A)_{11} \approx \frac{1}{3} \cos^2 \beta$ , $(k_A)_{22} \approx  \frac{1}{3} \sin^2 \beta$ , $(k_A)_{33} \approx \frac{1}{3} \sin^2 \beta$, with $\tan\beta \in [0.25,\,170]$ the ratio between the vacuum expectation values of the up- and down-sector Higgs doublets \cite{\diluzio}. This implies
\bea
C_{ann}\vert_{ \rm DFSZ}&=& -0.123(30)+0.406(15)\sin^2(\beta)~,\\
C_{app}\vert_{\rm DFSZ}&=&-0.169(30)-0.430(15)\sin^2(\beta)~.
\eea
\item ``Agnostic'': in this case the axion-nucleon couplings are solely required to comply with existing data. Measurements of Neutron Star (NS) cooling yield  $C_{ann}\,m_N/f_a$ $\lesssim 1.15 \times 10^{-9}$ and $C_{app}\,m_N/f_a \lesssim 1.09 \times 10^{-9}$~\cite{Buschmann:2021juv}. We will refer to these limits as the NS bounds.
In addition, we apply SN cooling constraints via the Raffelt bound~\cite{Raffelt:1990yz,Burrows:1988ah} $L_a \lesssim L_\nu$, with the neutrino luminosity $L_\nu \close 3\times 10^{52} \mathrm{erg}/\mathrm{s}$~\cite{Burrows:2000mk,Woosley:2005cha} inferred from SN~1987A~\cite{Bionta:1987qt,IMB:1988suc,Kamiokande-II:1987idp,Hirata:1988ad,Alekseev:1987ej,Alekseev:1988gp}. These constraints are henceforth referred to simply as SN bounds. The procedure to calculate them is detailed around \refeq{eq:La}.

In the numerics we take $C_{ann}$, $C_{app}$ as large in magnitude as the smallest between the NS and the SN bounds.
\end{itemize}

\noi The remaining Lagrangian terms required read
\be
\mc L_{N N \gamma} = - e \, \bar p \, \slashed A \,  p~,
\ee
\be
\mc L_{NN' \pi} = \frac{D+F}{\sqrt2} \left( \frac{\de_\mu \pi^0}{\sqrt2 F_0} \Bigl( \bar p \gamma^\mu \gamma^5 p - \bar n \gamma^\mu \gamma^5 n\Bigl) + \Bigl( \frac{\de_\mu \pi^+}{F_0} \bar p \gamma^\mu \gamma^5 n + \hc \Bigl) \right)~.
\ee
In addition, $\mc L_{a \pi \pi'} = 0$, hence so is $\mc L_{a \pi \pi^0}$. As a consequence, the vertex induced by the contraction of the $\pi$ in these two interactions is absent, unlike the $SU(3)$ case \cite{\mael}. Finally, the 4-leg vertex reads
\be
\mc L_{a \pi N N^{\prime}} = -i \frac{\de_\mu a}{f_a} \, C_{a\pi N} \dfrac{\pi^-}{F_0} \bar n \gamma^\mu p+ \hc~,
\ee
where
\bea
	C_{a\pi N}&=& \dfrac{\cgg}{\sqrt{2}} \dfrac{1-z}{1+z}+\dfrac{(k_A)_{11}-(k_A)_{22}}{2\sqrt{2}}~.
\eea
The full parametric dependence of our processes of interest can be studied in the $C_{ann}$ vs. $C_{app}$ plane because
\be
	C_{a\pi N}=\dfrac{C_{app}-C_{ann}}{2\sqrt{2}g_A}~,
\ee
where $g_A=-(D+F)=1.26(5)$. 

The above interactions induce our processes of interest, to be discussed in the next section, via axion interactions with hadrons. One may also include the contributions due to the $a\gamma\gamma$ interaction
\be
\label{eq:Layy}
\mc L_a \supset c_{a \gamma\gamma} \frac{\alpha}{4 \pi} \frac{a}{f_a} F_{\mu \nu} \tilde F^{\mu \nu}~,
\ee
where $\tilde F^{\mu \nu} \equiv \frac{1}{2}\epsilon^{\mu \nu \rho \sigma} F_{\rho \sigma}$.
However, unless axion-nucleon couplings are many orders of magnitude below the values allowed by current data, we generally expect the axion-nucleon interactions to be dominant with respect to the interaction \refeq{eq:Layy}, which comes with a relative $\alpha / (4 \pi)$ suppression at amplitude level.

\section{Detection on Earth of Supernova Cooling Particles}\label{sec:detection}

The core collapse and cooling that drives the transition from a SN to a young neutron star produces a burst of neutrinos and may similarly emit fluxes of additional feebly interacting, uncharged particles such as axions.
The flux of these particles may be detected on Earth by exploiting in reverse sense their interaction with nucleons, producing \che light. Possible processes include
\bea
\label{eq:processes}
\bar \nu \, p ~\to~ n \, e^+ ~,~~~~~
\nu \, n ~\to~ p \, e^- ~,~~~~~
a \, p ~\to~ p \, \gamma ~,~~~~~
a N \to N \pi^0~,~~~~~
\eea
with $N$ either of $n, p$ in the last process and henceforth where this symbol is used.
All these processes are of the form
\be
\label{eq:XN_Nlight}
\axnu \, N ~\to~ N^{(\prime)} \,\chep~, 
\ee
where the $\axnu$ particle is absorbed by a nucleon $N$, to yield a final-state nucleon $N^{(\prime)}$ and a ``\che parton'' $\chep$ ($= e^\pm, \gamma, \pi^0$), that initiates an electromagnetic cascade eventually detected as \che light.
The total cross section $\sigma_{\axnu\chep}$ and the number spectrum $d N_\axnu / d E_\axnu$ allow to calculate the number spectrum of the particle $\chep$ as%
\footnote{\refeq{eq:dN_chep} is not necessarily obvious. However, it follows in a straightforward way from writing the total $N_\chep^{(\axnu)}$ as
\be
N_\chep^{(\axnu)} = \int_{m_\chep}^{+\infty} d E_\chep \frac{d}{d E_\chep}\left(\int_{m_\axnu}^{+\infty} d E_\axnu \frac{N_t \sigma\left(E_\axnu\right)}{4 \pi d^2} \frac{d N_\axnu}{d E_\axnu}\left(E_\axnu\right)\right)~.
\ee
In full generality, the integration should be carried out over the phase space of the initial nucleon. However, in the laboratory frame, we approximate $\frac{dN_N}{dE_N}(E_N) = N_t \delta(E_N - m_N)$ because $T_{\text{water}} \ll m_N$.}
\be
\label{eq:dN_chep}
\frac{d N^{(\axnu)}_\chep \left( E_\chep \right)}{d E_\chep} = \frac{N_t}{4 \pi d^2} \int_{m_\axnu}^{+\infty} d E_\axnu \, \frac{\de \sigma_{\axnu\chep} \left(E_\chep, E_\axnu\right)}{\de E_\chep} \, \frac{d N_\axnu \left(E_\axnu\right)}{d E_\axnu}~.
\ee
In this relation, $d$ is the SN-to-Earth distance and $N_t$ is the number of targets in the detector. The case of HK corresponds to $N_t = 10^9 \times N_p \times N_A \times (M_{\rm det} / {\rm kton}) \times (g/{\rm mol} / m_{\aqua})$, with $N_p = 2$ the number of protons per water molecule and $N_A$ the Avogadro number.

$E_\chep$ can be expressed as a function of $E_\axnu$, of the external-state masses $m_{N,N^\prime,\axnu,\chep}$, and of $\cos \theta_{\axnu\chep}$. A useful limit in the present context is when $m_{N^\prime} = m_N$ and $m_{\axnu,\chep} \ll m_N$. In this limit, the relation becomes\footnote{In our case, $m_\chep \le m_{\pi}$, with equality attained when the \che parton is a $\pi^0$. Note also that corrections to this limit of the kinematic equation are quadratic in the ratios $m_{\axnu,\chep} / m_N$.}
\be
E_\chep = \frac{|\vec{p}_\axnu|}{1 + \frac{|\vec{p}_\axnu|}{m_{N}} \left(1 - \cos \theta_{\axnu\chep} \right)}~.
\ee
If the \che parton is a photon, and in the limit ${|\vec{p}_\axnu|}/{m{N}} \to 0$, one finds that $E_\chep = E_\axnu$, as in Ref.\cite{\lucente}. Similarly, \refeq{eq:dN_chep} reduces to the expression used in Ref.~\cite{\lucente} in this limit.

\begin{figure}
\centering
\includegraphics[width=0.3\linewidth]{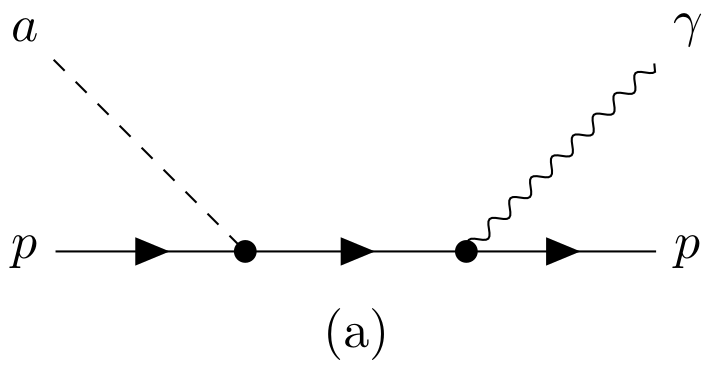}~~
\includegraphics[width=0.3\linewidth]{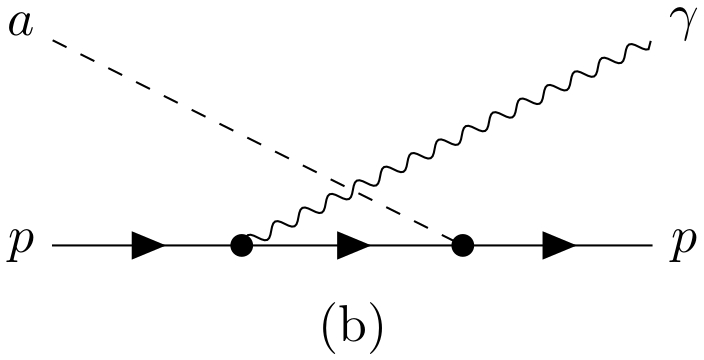}~~
\includegraphics[width=0.27\linewidth]{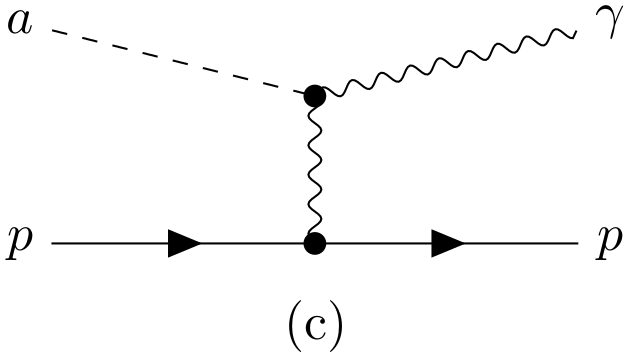}
\includegraphics[width=0.3\linewidth]{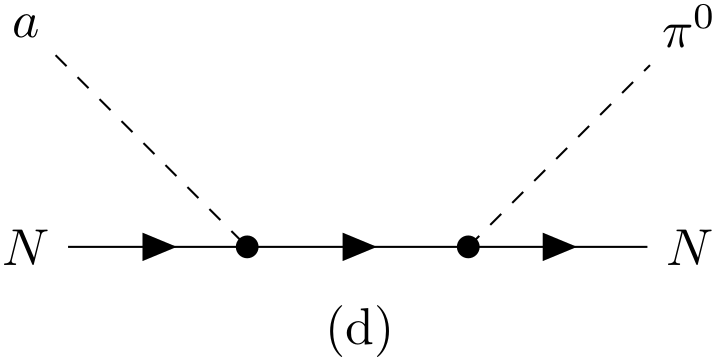}~~
\includegraphics[width=0.3\linewidth]{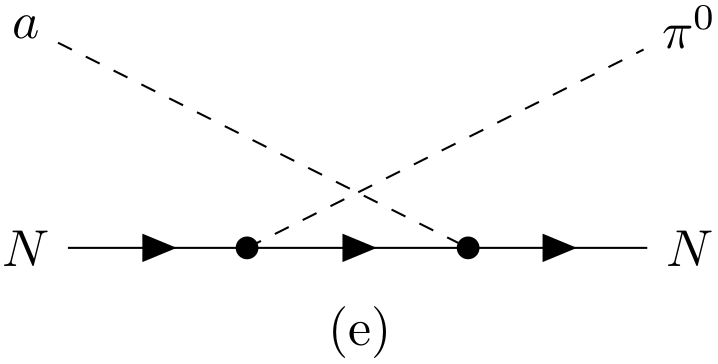}
\caption{Diagrams contributing to photon or $\pi^0$ emission from axion absorption by nucleons.} 
\label{fig:aN_to_gamma}
\end{figure}

We next discuss in detail the two key quantities entering \refeq{eq:dN_chep}, namely the cross section $\sigma_{\axnu\chep}$ and $N_\axnu$. In the case $\axnu = \nu$ we take these quantities from Ref.~\cite{Fogli:2004ff}. Alternatively, the neutrino number spectrum may be obtained from the time-integrated spectrum, normalized to the total number of events expected at HK. The obtained shape is consistent with the results in Ref.~\cite{\lucente}. 
We thus discuss in detail the axion case only ($\alpha = a$). Consideration of the cross section $\sigma_{a\chep}$ assumes that the axion interacts with {\em free} nucleons in water, which is a good approximation if we restrict to the two hydrogen atoms in each water molecule~\cite{\lucente}.
The $\sigma_{a\chep}$ cross section is then readily calculable, and the diagrams leading to the last two processes in \refeq{eq:processes} are shown in \reffig{fig:aN_to_gamma}.
Explicit expressions for the squared amplitudes necessary for the computation of $\sigma_{aX}$ ($X = \gamma$ or $\pi^0$) are collected in \refapp{app:sigma}.
The contribution from diagram~(c), induced by the interaction \refeq{eq:Layy}, has been commented upon at the end of \refsec{sec:theory_setup} and will not be considered further.

We now turn to the calculation of the differential number spectrum $dN_a / dE_a$ of the axion emitted from the SN. We start from the number density spectrum per unit time, given by
\renewcommand{\axnu}{a}
\bea
\label{eq:dn}
\frac{d \dot n_\axnu}{d E_\axnu} ~=~ \int (2\pi)^4 \delta^4(p) \, \left\vert \mc M \right\vert^2 \df_i \, \df_M \, (1 - \df_f) \, \frac{d \Omega_\axnu E_\axnu}{2 (2\pi)^3} \, \prod_{m}^{\{i,f,M\}} \frac{d^3 \vb*{p}_m}{(2\pi)^3 2 E_m}~,
\eea
where $\left\vert \mc M \right\vert^2$ denotes the absolute value squared of the amplitude $\sum_n \mc A(\Bimba)$, $n = \{ i,f,M\}$, and the indices $i, f, M$ label the different possible instances of initial-, final-state baryon, and meson involved in the reactions. The hadron distribution functions in the medium are defined as 
\be
\label{eq:pdf}
\df_{j} ~=~ \frac{1}{\exp \left( \frac{E_{j} - \mu_{j}}{T} \right) + (-1)^{2 s_j + 1}}~,
\ee
where $s_j$ denotes the spin of the corresponding particle. Contributions from decays of the kind $\Biba$ can be accounted by modifying \refeq{eq:dn} in an obvious way.

Three important remarks are in order. The first concerns the processes to include in \refeq{eq:dn}. As the relation is written, all processes of the kind $\Bimba$ as well as $\Biba$ would have to be included, where $B_{i,f}$ and $M$ denote respectively octet baryons and octet mesons. Such calculation has been performed in Ref.~\cite{\mael}, with a phase space corresponding to the SN~1987A event \cite{Bionta:1987qt,IMB:1988suc,Kamiokande-II:1987idp,Hirata:1988ad,Alekseev:1987ej,Alekseev:1988gp}. However, for consistency with the effective theory used for calculating axion absorption in water \che, we only keep the processes where $B_{i,f}$ are nucleons and $M$ are pions. Second, even focusing on nucleons only, one should in principle include reactions that radiate axions in nucleon-nucleon scattering, $N_1 N_2 \to N_3 N_4 \, a$, as they provide the most reliable bound~\cite{Caputo:2024oqc}. However, as discussed in~\cite{\carenza}, the axion number spectrum produced via these channels peaks at a clearly separated, and {\em lower}, energy than the $N \pi \to N^\prime a$ processes we restrict our attention to. Including the contribution from the high-energy tail of $N_1 N_2 \to N_3 N_4 \, a$ processes will only strengthen our conclusions.
Third, \refeq{eq:dn} treats all particles involved as ideal gases. However, in-medium effects in hot and dense matter are expected to have a strong impact~\cite{Martinez-Pinedo:2012eaj,Roberts:2012um}. One way to capture them is by following the approach in Ref.~\cite{\mael}, i.e. to specify the necessary input using different benchmark models for the thermodynamic conditions inside the SN as well as for the EOS determining all the particles fractions. We survey this dependence by considering \DDmod~\cite{\ddref} and \SFmod~\cite{\sfref} as alternative EOS models, as well as two possible temperatures for the SN core, $T = \{ 30, 40\} ~\MeV$, and two possible values for the baryon number density $n_B$, namely $\{ 1, 1.5\} \, n_\sat$, where $n_{\sat} = 1.6 \times 10^{38} \, \cm^{-3}$. This approach is entirely calculable if we further specify a third thermodynamical parameter, such as the proton fraction normalized to the total baryon fraction, $Y_p$, that we fix to the value 0.3 (e.g. \cite{\carenza}).

From \refeq{eq:dn}, the particle number spectrum can be obtained integrating over time and over the volume of the emitting object.\footnote{This assumes that the axion is in the free-streaming regime.
%Similarly as Ref.~\cite{\mael}, we verified that this assumption holds in the full parametric space of interest for this work.
%We verified that this assumption holds in the full parametric space of interest for this work. This result is unsurprising, because the SN models considered here are very close to those considered in Ref.~\cite{\mael}, and the parametric space of interest in this work implies axion couplings to matter as weak as, or weaker than, those considered in Ref.~\cite{\mael}.
We verified that this assumption holds across the entire parameter space relevant to this study. This outcome is expected, as the SN models adopted here are very close to those in Ref.\cite{\mael}, and the parameter space considered entails axion couplings to matter that are at least as weak as, if not weaker than, those in Ref.\cite{\mael}.
}
Similarly as much of the existing literature, we assume \refeq{eq:dn} to be constant over an emission time interval $\Delta t \close 10s$, and also over the SN volume, estimated from $M_{\rm SN} / \rho_0$, with $\rho_0 \close 2.6 \times 10^{14}~{\rm g}/{\rm cm}^3$ the nuclear saturation density. Also, following the classic argument in Ref.~\cite{Raffelt:1990yz,Raffelt:2006cw} (see also the recent Ref.~\cite{Caputo:2024oqc}), we assume $M_{\rm SN}$ to equal one solar mass. A working estimate of the axion number spectrum may thus be obtained from
\be
\label{eq:dN_vs_dn}
\frac{d N_\axnu}{dE_\axnu} ~=~ \kappa ~ \frac{d \dot{n}_\axnu}{dE_\axnu}~,
\ee
with $\kappa = (\Delta t) \times M_{\rm SN} / \rho_0$. In our numerical analysis, we stick with this approach. From \refeq{eq:dN_vs_dn} we can estimate the luminosity radiated in axions via~\cite{Raffelt:1990yz,Raffelt:2006cw}
\be
\label{eq:La}
L_a\Delta t ~=~ \int_{m_a}^{+\infty}dE_a E_a \dfrac{dN_a}{dE_a}~,
\ee
with $\Delta t \close 10s$.\footnote{In the numerics, $\sim$ is replaced with $=$. The $\sim$ symbol is used to indicate inherently crude estimates.} This allows to implement the Raffelt bound relevant for the ``agnostic'' case of interest mentioned on page \pageref{page:models}. The integral in \refeq{eq:La}, and thus the SN constraint, depends on the choice of the SN model, which comprises the EOS and the thermodynamical parameters. The upper bounds on $C_{ann}$, $C_{app}$ inferred from \refeq{eq:La} for the different SN models we considered are tabulated in Appendix~\ref{app:bounds}.

\newcommand{\redshi}{\xi}
\renewcommand{\axnu}{\alpha}
The actual value of $\kappa$ is subject to significant uncertainty, primarily due to the limited knowledge of the emitter. We outline the uncertainties we are aware of in the next paragraph. Importantly however, none of them is expected to alter our main conclusions. 
In fact, in our study we focus on locating and comparing the peak positions and heights in the spectra for different \che partons $\chep$, depending on whether $\axnu$ is taken as a neutrino or an axion. Our aim is to determine if, under typical parameter choices for SN emission on one side, and detection in water tanks on the other side, the axion-originated spectrum can be distinguished from the neutrino spectrum. The effects enumerated above will be common to {\em all} the spectra we are considering. Hence these effects, along with a detailed evaluation of the relationship in \refeq{eq:dN_vs_dn}, will modify the spectra similarly across different $\chep$ types, likely without altering the peak positions or order-of-magnitude heights. Therefore, we could in principle restrict our focus to ${d \dot{n}_\chep}/{dE_\chep}$. Our choice of $\kappa$ primarily serves to provide the spectrum with correct mass dimensions and facilitate comparisons with existing literature.

Having made the above qualification, we now enumerate the most important sources of uncertainty that we necessarily neglect. First, our discussion concerns SN-like events that have yet to be observed, and the quantities involved in predicting spectra vary across different emitting bodies. Second, even within an identified SN event, neither the emitter’s mass, the precise emission time, nor the exact fraction of mass lost to radiation is known to better than order-of-magnitude accuracy. Third, inhomogeneities within the $\Delta t$ emission interval and spatial irregularities, such as radial density variations or deviations from spherical symmetry~\cite{Fischer:2021jfm, Mori:2021pcv, Betranhandy:2022bvr, Mori:2023mjw}, may also affect the results. Fourth, the flux that travels the distance from the SN to Earth must be corrected for gravitational and observer-time redshifts, each encoded in a factor $\redshi$, as well as for the Doppler-shift effect due to the radial velocity $v_r$ of the medium. These effects are subsumed in the multiplicative correction $\redshi^2(r)(1 + 2 v_r)$ \cite{\caravi,\cajaravi}, which collectively reduces the luminosity by approximately 20–30\% \cite{\caravi}.

We reiterate, however, that all these known unknowns are not expected to alter our conclusions, as argued below \refeq{eq:La}.

\section{\che spectra} \label{sec:spectra}

\noi The previous discussion allows to calculate the \che-parton number spectrum, \refeq{eq:dN_chep}, where the $\axnu$ particle absorbed in the \che-light  producing process, \refeq{eq:XN_Nlight}, is an axion.

\subsection{Axion- vs. neutrino-absorption spectra}

\subsubsection{Physical spectra}

\begin{figure}[t]
\centering
\includegraphics[width=0.49\linewidth]{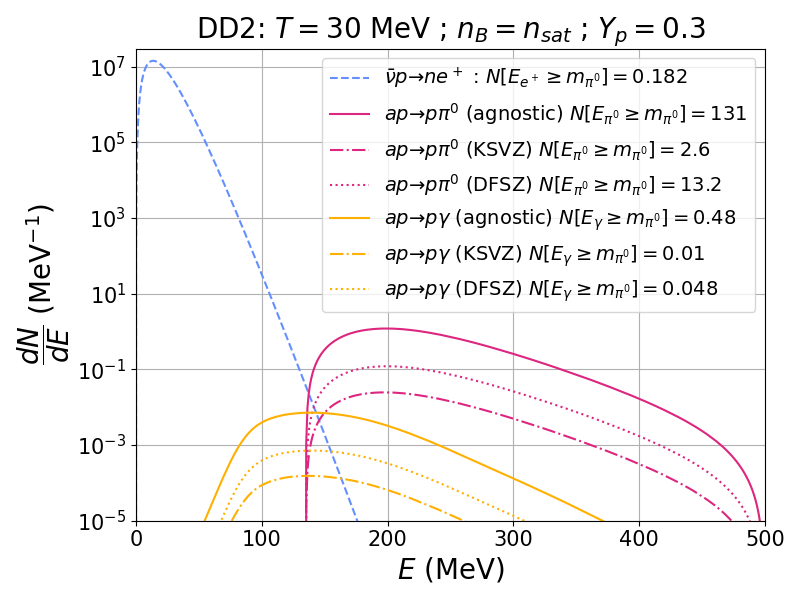}
\includegraphics[width=0.49\linewidth]{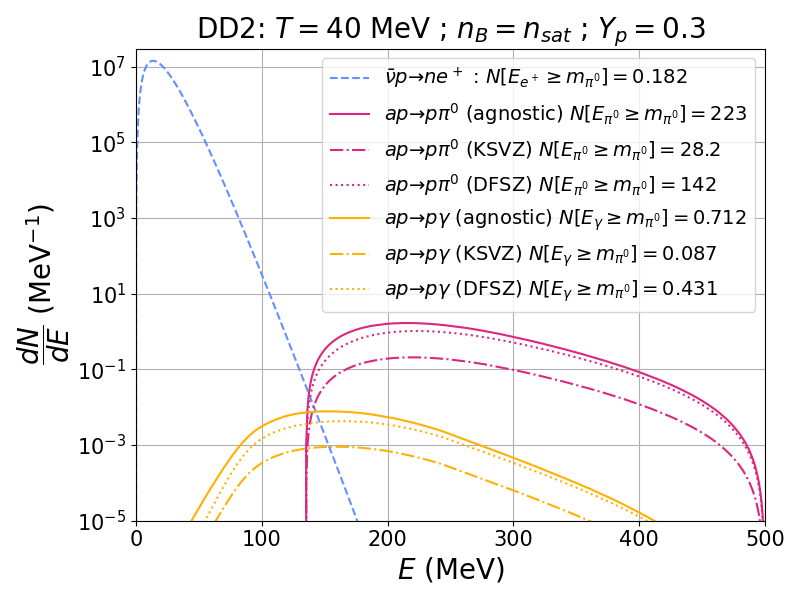}
\includegraphics[width=0.49\linewidth]{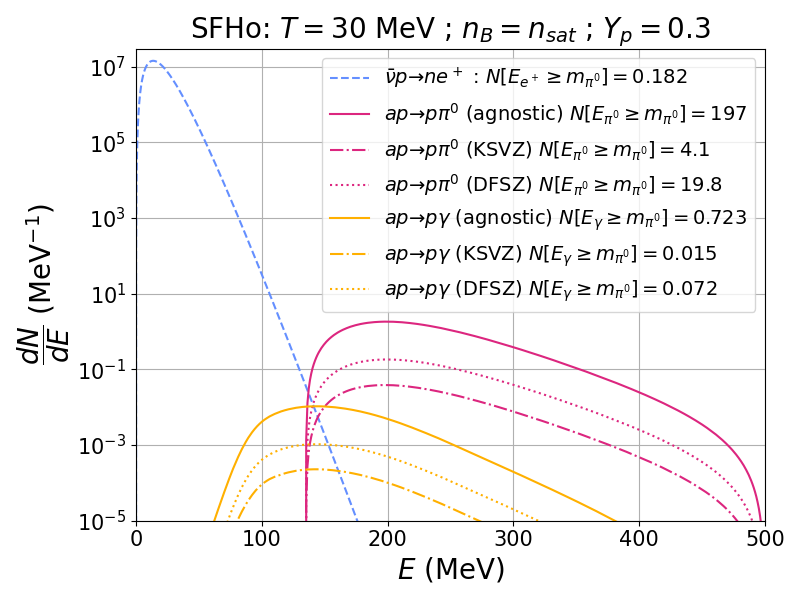}
\includegraphics[width=0.49\linewidth]{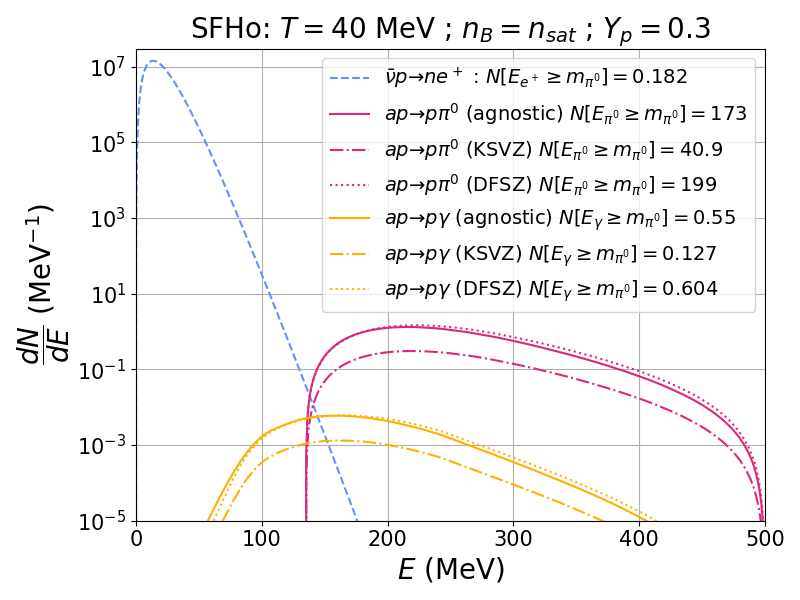}
\caption{Comparison between the \che-light spectra induced by neutrino (dashed blue) vs. axion absorption. The latter includes either a photon (yellow) or a $\pi^0$ (purple) as \che partons, and different line styles denote the different theory cases discussed in \refsec{sec:theory_setup}---see legend, which specifies also the total number of events above the $\pi^0$ threshold. The first (second) row refers to the \DDmod (\SFmod) EOS models, and the first (second) column to $T = 30$ (40) $\MeV$. The remaining thermodynamic parameters are fixed to a reference value, as their variations within motivated ranges are imperceptible.}
\label{fig:money}
\end{figure}

Following the discussion leading to \refeq{eq:dN_vs_dn}, in the plots of \reffig{fig:money} we compare the calculated axion and neutrino number spectra.
The two rows refer to the \DDmod and \SFmod SN models, respectively, and the left vs. right columns use $T = 30$ vs. $40~\MeV$. All plots use $n_B = n_\sat$, and we omit to show the $n_B = 1.5\, n_\sat$ counterparts as differences are not significant.

Each of the panels in \reffig{fig:money} displays {\em three} groups of number spectra for the different \che partons:
\begin{itemize}
\item in dashed blue, the $e^+$ number spectrum from the $\bar \nu p \to n e^+$ process \cite{Fogli:2004ff}, peaking leftmost in all panels;
\item in purple, and with different line styles according to the theory framework used, the spectrum of the $\pi^0$ in $a p \to p \pi^0$, peaking rightmost in all panels;
\item in yellow, and again with different line styles, the spectrum of the photon in $a p \to p \gamma$, the smallest-peaking in all panels.
\end{itemize}
The spectra for the axion-absorbing processes depend on the values of the underlying axion-nucleon couplings $C_{ann, app}$, which are fixed according to the three alternative cases of interest outlined in the itemization on page \pageref{page:models}. These cases are shown as different line styles (solid, dot-dashed, dotted), as detailed in the legend.

We also display, again in the legend, the integral of each spectrum for $E_X \ge m_{\pi}$ ($X$ denoting the relevant \che parton). We made this definite choice for the lower boundary because, as it can be seen from all panels, it is a simple proxy of the point at which the spectra for the $\bar \nu p \to n e^+$ and the $a p \to p \pi^0$ processes intersect. The numbers in the legend suggest that the $\pi^0$ process will induce hundreds of events for energies above $m_\pi$, with basically no background from the $\bar \nu p \to n e^+$ process. These numbers also indicate that the $a p \to p \gamma$ process contributes at most $O(1)$ events in this region, and is thus much less promising in comparison with the $\pi^0$ process. These conclusions are expected to hold even if we were able to include all the effects discussed in the concluding paragraph of \refsec{sec:detection}.

It is worth noting that the number of expected events increases with temperature in the KSVZ and DFSZ scenarios, as expected. However, this is not necessarily the case when only experimental bounds are imposed (the ``agnostic'' case). For example, within the \SFmod model, an {\em increase} in temperature causes the Raffelt bound to be saturated at {\em lower} values of $C_{aNN}$, which in turn leads to a lower number of events.

\subsubsection{Detector-response corrected spectra}

In this section, we assess the expected reconstruction efficiency for the $\pi^0$ spectra in \reffig{fig:money}, based on a realistic simulation of the detector response. The underlying motivation for this line of inquiry is the fact that, unlike positrons emitted from inverse-$\beta$ decay of SN antineutrinos, $\pi^0$s, being electromagnetically neutral, do not directly produce \che radiation and are instead detected via their e.m. decay into two high-energy photons. The latter provide a distinct signature with respect to the \che cascade initiated by the $e^+$ originating from antineutrino absorption. Furthermore, if {\em both} photons are reconstructed, it is possible to calculate the invariant mass of the $\pi^0$, adding additional validation to the signal. The study in this section provides the context towards addressing these two questions in an actual experimental analysis.

To estimate the reconstruction performance and signal detection at HK, we simulate the $\pi^0$ decays using the {\tt PYTHIA8} software~\cite{Bierlich:2022pfr} and adapt the existing and validated Monte-Carlo simulations and tools of atmospheric neutrinos~\cite{DVN/OS5N7U_2023}.
Since the $\bar \nu$ signal lies just at the 100~MeV threshold for atmospheric neutrino analyses\cite{hk_tdr, Wester_2024}, we extend the detector response to lower energies and relax the energy cuts applied in these analyses. This adjustment improves the efficiency of the \textit{2-ring $\pi^0$-like} sample. Furthermore, given that the expected signal is confined to a few seconds, this modification has virtually no impact on contamination from lower-energy events or accidental coincidences with cosmic rays.

According to the sample definitions, there are four different topologies in which these events can be classified:
\begin{itemize}
    \item \textit{2-ring $\pi^0$-like}: Two photons are reconstructed, and their invariant mass falls between 85~MeV and 215~MeV.
    \item \textit{1-ring $\pi^0$-like}: A single \che cascade is reconstructed and satisfies specific selection criteria based on likelihoods computed under the assumption that it originates from a $\pi^0$ decay with a small opening angle between the two gammas.
    \item \textit{1-ring e-like}: A single \che cascade is reconstructed but is found to be incompatible with the $\pi^0$ decay hypothesis.
    \item \textit{1-ring $\mu$-like}: A single \che ring is reconstructed and identified as compatible with a muon signal. This sample primarily consists of events with misidentified particle types.
    \item \textit{2-ring other}: Additionally, we consider events with two reconstructed photons that do not satisfy the $\pi^0$-like criteria. These events correspond to poorly reconstructed $\pi^0$ decays that remain distinguishable from the neutrino signal.
\end{itemize}

\begin{figure}
    \centering
    \includegraphics[width=0.95\linewidth]{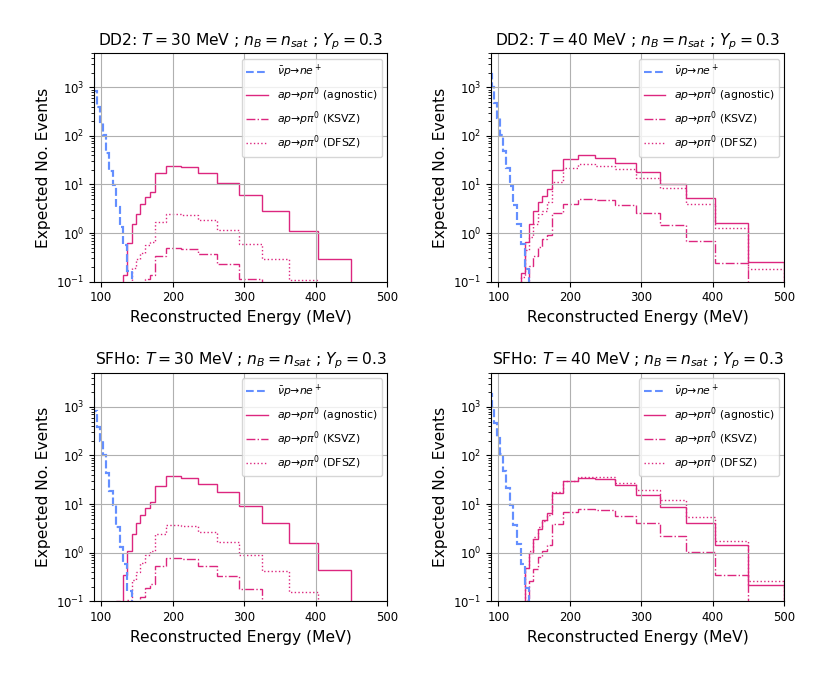}
    \caption{Reconstructed energy for antineutrino inverse-$\beta$ decay (blue) and axion-induced $\pi^0$ emission (purple) with different line styles denoting the different theory cases discussed in \refsec{sec:theory_setup}.}
    \label{fig:ereco_models}
\end{figure}

\reffig{fig:ereco_models} shows the reconstructed energy spectra of neutrino- and axion-induced signals for the different models assumed in \reffig{fig:money}. Motivated by this figure and the considerations around it, an experimental energy cut at 150~MeV would isolate the $\pi^0$ signal. 

\reftab{tab:reco_T_30_40} shows the expected efficiency and purity of each sample. The energy cut is sufficient to reduce drastically the contamination from the neutrino emission. Most of the $\pi^0$ events fall under the \textit{2-ring $\pi^0$-like} category, as expected from lowering the detection threshold in the simulations and due to the low energy of the neutral pions, which implies large opening angles between the decay gammas and enables a good separation and reconstruction. The events with poorer reconstruction populate the 1-ring samples, with a small fraction also classified as \textit{2-ring other}, in which at least one of the \che rings is misidentified. Any other contribution to the event rate, such as atmospheric neutrinos, is expected to be negligible due to the reduced time window of the signal.\footnote{For completeness, in \refapp{app:T_30_40} we further split each event category considered in \reftab{tab:reco_T_30_40} according to 
the two SN models considered, \DDmod and \SFmod, as well as the three alternative ways---agnostic, KSVZ, and DFSZ---of constraining the relevant axion interactions.}

\reftab{tab:reco_T_30_40} confirms that the $\pi^0$ signal overwhelmingly dominates for reconstructed energies above 150~MeV. More importantly, they indicate that the $\pi^0$ reconstruction efficiency is expected to be high, around 80\%, though not perfect, as about 20\% of the rings will not exhibit a $\pi^0$-like shape.

The signal above 150~MeV is essentially background-free. Thus, accurately assessing the reconstruction efficiency will be relevant when distinguishing between axion-induced signals and other potential physical sources such as those considered in Refs.~\cite{Mastrototaro:2019vug,Fiorillo:2022cdq}, whose spectra partially overlap with our signal.\footnote{We thank P.D. Serpico for this comment.}

\newcommand{\reco}{{\rm reco}}
\begin{table}[h]
    \centering
    \begin{tabular}{lcc}
        \toprule
        & \multicolumn{2}{c}{\textbf{Reconstructed as} $\bm{\pi^0}$} \vspace{0.1cm} \\
        \textbf{Selection} & $\bm{T = 30~\MeV}$ & $\bm{T = 40~\MeV}$ \\
        \midrule
        $E_{\reco} > 150~\MeV$ & 0.973 & 0.984 \\
        $E_{\reco} > 150~\MeV$, 2-ring $\pi^0$-like & 0.675 & 0.582 \\
        $E_{\reco} > 150~\MeV$, 1-ring $\pi^0$-like & 0.146 & 0.201 \\
        $E_{\reco} > 150~\MeV$, 1-ring $e$-like & 0.084 & 0.134 \\
        $E_{\reco} > 150~\MeV$, 1-ring $\mu$-like & 0.007 & 0.013 \\
        $E_{\reco} > 150~\MeV$, 2-ring other & 0.061 & 0.053 \\
        \bottomrule
    \end{tabular}
    \caption{Fractions of axion-induced \che events reconstructed at HK, categorized according to the event types discussed in the main text. Axion emission follows the SN \DDmod model with $T = 30$ or $40~\MeV$. The corresponding fractions of $\bar{\nu}$-induced \che events are omitted as they are always smaller than $O(10^{-11})$.}
    \label{tab:reco_T_30_40}
\end{table}

\subsection{Number of events as a function of the axion-matter couplings}

Next, focusing on axion absorption, \refeq{eq:dN_chep} allows to calculate the expected number of signal events $\left<N_X^{(R)}\right>$. The total number spectrum is the result of an integral over a specified energy range $R$. Given that the $dN_X$ spectra of the \che partons induced by the neutrino ($X = e^+$) vs. the axion ($X = \gamma$ or $\pi^0$) peak at well-separated energies, of order 10~MeV for the $e^+$ and 100-200~MeV for the $\gamma$ or the $\pi^0$, in general one may define the energy range $R~\equiv~ [ E^*, \infty )$, with $E^*$ the energy value above which $dN_X^{(a)}/dE_X > dN_{e^+}^{(\nu)}/ dE_{e^+}$. However, from the previous section, we saw that $X = \pi^0$ is the vastly dominant component, and for this component we can take $E^* \simeq m_{\pi}$, so we will henceforth restrict our numerical analysis to this component, taking $R  = [m_{\pi}, \infty )$.

Then the expected number of signal events may be estimated as
\be
\label{eq:lambdaR}
\left<N_{\pi^0}^{(R)}\right> ~\equiv~ \lambdaR(C_{app}, C_{ann}) ~=~ \int_R dE_{\pi^0} \frac{dN_{\pi^0}^{(a)} (C_{app}, C_{ann})}{dE_{\pi^0}}~.
\ee
The probability of observing any given number $n$ of actual events obeys a Poisson distribution with parameter~$\lambdaR$, $P(n; \lambdaR)$. The observed number of events can thus be estimated as $\lambdaR \pm \delta(p)$, where $\delta(p)$ depends on the considered $p$-value
\be
\label{eq:deltap}
\sum_{k \in [\lambdaR - \delta (p),~\lambdaR + \delta (p)]} P(k; \lambdaR) ~=~ p~.
\ee
An equivalent way of estimating this interval is by defining the Poisson-distributed test statistic $t_P(n,\lambdaR) \equiv (n - \lambdaR)^2 / \lambdaR$ \cite{\pdg}. If $\lambdaR \gg 1$, $t_P$ follows a $\chi^2$ distribution with $N_{\dof} = 1$, so that e.g. $\chi^2(p) / N_{\dof} = \{ 1, 9 \}$ for $68.3$ and respectively $99.7\%$ probability. From the equality $t_P = \chi^2(p) / N_{\dof}$ one can then determine $\delta(p) = \sqrt{\lambdaR \, \chi^2(p) / N_{\dof}}$.
With a slight abuse of language, we will denote $\lambdaR \pm \sqrt{\lambdaR}$ simply as the 1$\sigma$ interval.\footnote{As discussed, this is well motivated for large $\lambdaR$. For small $\lambdaR$, we have numerically verified that $\lambdaR \pm \sqrt{\lambdaR}$ corresponds to a 68.3\% probability.} This interval depends on the SN modeling (see discussion below \refeq{eq:pdf}), as the latter affects the axion flux on Earth.

\begin{figure}
\centering
\includegraphics[width=\linewidth]{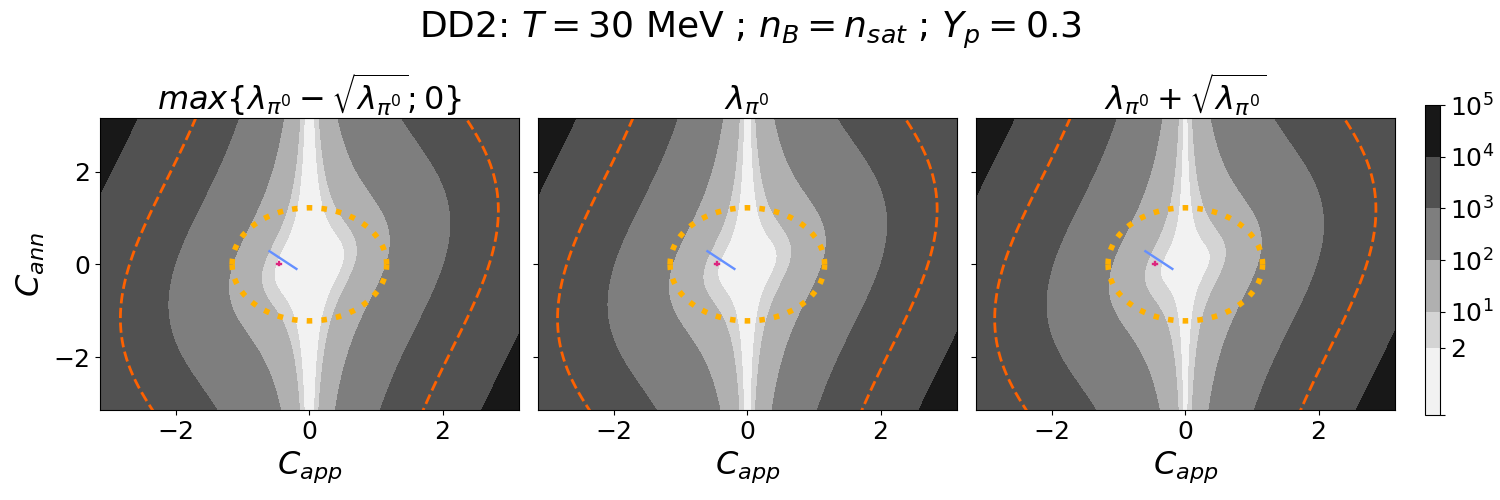}
\includegraphics[width=\linewidth]{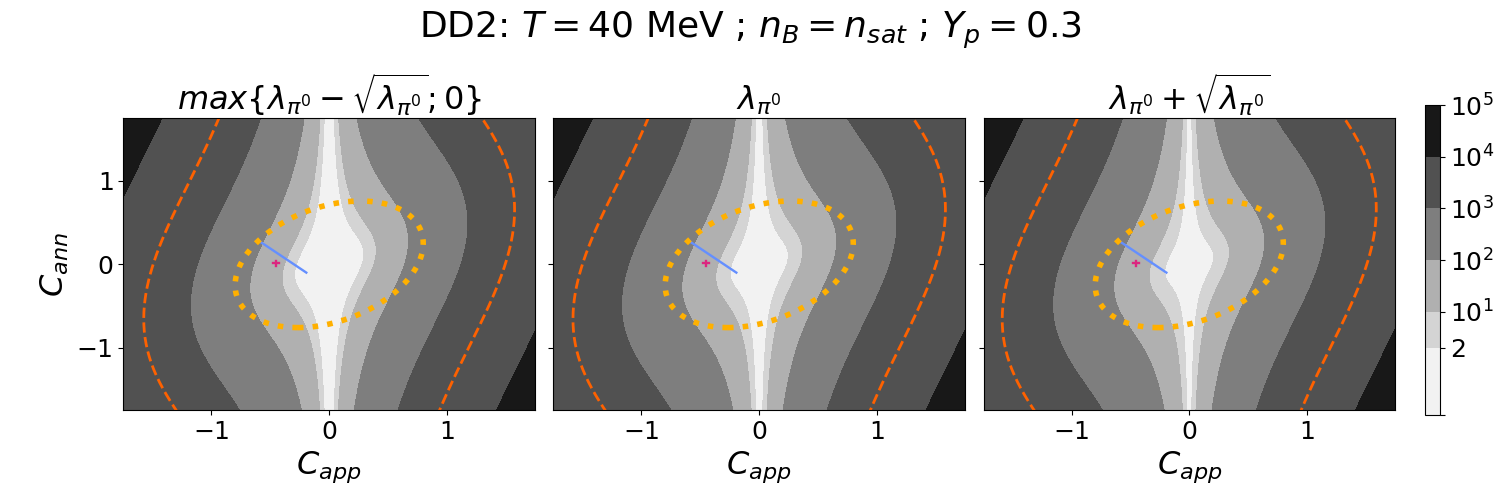}
\caption{
Expected number of events (shades of grey) in the $C_{app}$ vs. $C_{ann}$ plane. The blue segment (DFSZ) extends outwards with $\tan\beta \in [0.25, 170]$, while the red cross represents KSVZ with uncertainties. The dotted yellow ellipses and the dashed orange lines indicate experimental constraints (NSs, SNe) and the detectability criterion ($N_{\pi^0} \geq 2$ for a Galactic center SN), respectively. The expected number of events is calculated as $\lambdaR$ in central panels, while left and right panels correspond to $\lambdaR \mp \sqrt{\lambdaR}$. The two rows represent $T = \{30, 40\}~\MeV$, respectively.
}
\label{fig:isobars_DD2_30_vs_40}
\end{figure}

In \reffig{fig:isobars_DD2_30_vs_40}, we show the expected number of events (represented as different shades of grey) in the plane $C_{app}$ vs. $C_{ann}$.
In each plot, the blue segment extending outward corresponds to DFSZ, as the value of $\tan\beta$ increases in the range $[0.25, 170]$, while the red cross represents KSVZ along with its uncertainties. The dotted yellow ellipses indicate the experimental constraints from NSs and SNe, while the dashed orange lines represent the detectability criterion $\lambda_{\pi^0}\geq 2$ for a candidate SN event at the Galactic center.

The number of events is calculated as $\laR$ (central panels) or $\laR \mp \sqrt{\laR}$ (left- or rightmost panels, respectively), showing barely perceptible differences. The two rows correspond to $T = \{ 30, 40\}~\MeV$, respectively. Once again, the upper and lower panels exhibit a very similar shape, but both the $x$- and $y$-axis scales change by a factor of about $1/2$. The dependence on the model choice (\DDmod vs. \SFmod) and on the $n_B$ value is not significant. Therefore, \reffig{fig:isobars_DD2_30_vs_40} conveys at a glance the full range of variability.

This variability can be summarized as follows. Regarding the $T$ dependence, which is by far the most pronounced, we note that moving from 30 to 40~MeV causes $n$ to change by a factor approaching 10.\footnote{This can be understood from the emissivity dependence: $Q_a \approx T^{7.5}$ multiplied by a factor with non-trivial $T$ dependence. If we neglect this factor, we would thus expect the temperature increase to yield approximately $8.7$ times more events.} For example, the KSVZ point yields about 5 events at $T=30~\MeV$ and approximately 50 events at $T=40~\MeV$. This demonstrates that the temperature dependence is quite pronounced. To provide conservative estimates, one may therefore stick to predictions at $T=30~\MeV$.

Concerning the dependence on the choice of $n_B$, all other parameters being equal, we find that $n_\sat$ vs. $1.5 \, n_\sat$ leads to an increase in the number of events by a factor between 1 and 1.3, with larger values for \SFmod than for \DDmod.
Finally, changing the model from \DDmod to \SFmod for a given value of $T$ increases the number of events by a factor between 1.5 (for $n_B = n_\sat$) and 1.8 (for $n_B = 1.5\, n_\sat$).

We emphasize that \reffig{fig:isobars_DD2_30_vs_40} and the above considerations are solely intended to provide an initial understanding of how the predicted number of events changes when varying reference SN parameters within fiducial ranges. A more robust analysis of this dependence would require comprehensive, state-of-the-art SN simulations, which are beyond the scope of the present work.

\subsection{Number of events vs. SN candidates}

\begin{figure}
\centering
\includegraphics[width=0.495\linewidth]{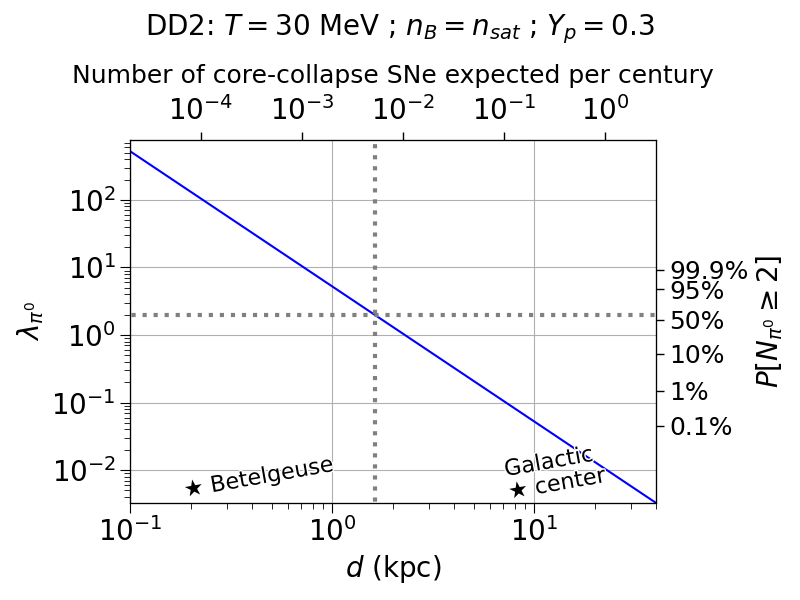}
\includegraphics[width=0.495\linewidth]{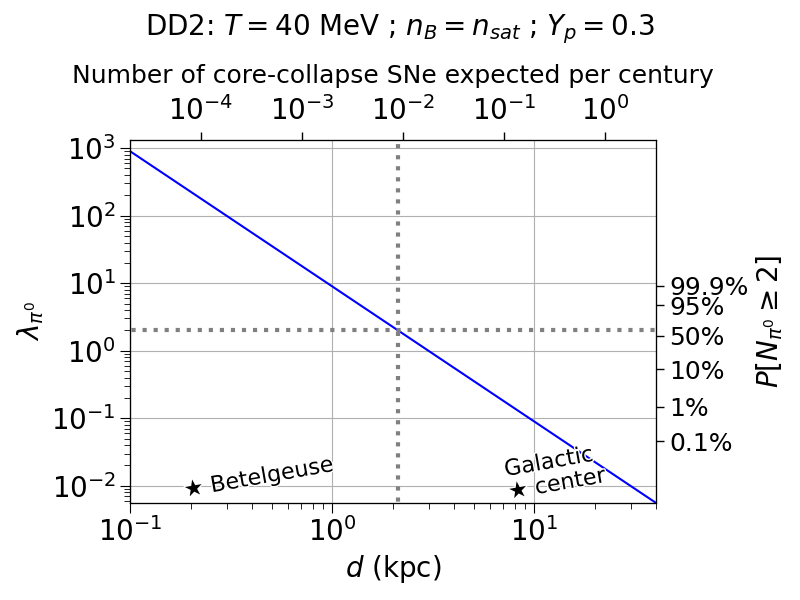}
\caption{Scaling of the expected number of $\pi^0$-initiated \che cascades with the inverse square of the distance $d$ of a SN candidate from Earth. The left $y$-axis shows the number of expected events as a function of $d$ (lower $x$-axis), with SN models matching those in \reffigs{fig:money}{fig:isobars_DD2_30_vs_40}. The grey dashed lines indicate the detectability threshold ($\geq 2$ events), defining the critical distance $d_2^{(a)}$ below which axion absorption remains observable. The upper $y$-axis maps $d$ onto the expected number of CC SNe per century, $\dot{N}(d)$, based on the estimation in Ref.~\cite{Tammann:1994ev}.
}
\label{fig:distance_plots}
\end{figure}
The calculated number of axion-absorption events for a given SN candidate scales quadratically with the inverse of the distance $d$ of the candidate from Earth, see \refeq{eq:dN_chep}. Beyond a certain distance, the number of signal events goes below 2, a number often taken as the ``detectability threshold''.\footnote{In \reffig{fig:money}, we can see that the expected number of events for $E \ge m_{\pi}$ from the neutrino-absorption process is 0.182 for $d=0.2$ kpc and accordingly smaller for larger distances. The $p$-value associated with observing 2 events for this expected number is 1.5\%.}
We display this scaling in \reffig{fig:distance_plots} in the case where the \che parton is the $\pi^0$, i.e. the last process in \refeq{eq:processes}. The SN models (see plot titles) are chosen to match the models in \reffigs{fig:money}{fig:isobars_DD2_30_vs_40}. The number of expected $\pi^0$-initiated \che cascades is shown on the left $y$-axis as a function of the distance $d$, shown in the lower $x$-axis, of the emitting candidate from Earth. The grey dashed lines crossing each other mark the distance, to be referred to as $d_2^{(a)}$, below which a SN candidate is above detectability threshold as far as {\em axion} absorption is concerned. The figure suggests that this distance is just below $2~\kpc$. Some important remarks should be made before proceeding further. First, $d_2^{(a)} \approx 2~\kpc$ refers to the detectability of the axion burst only---as well known, SNe are detectable from their neutrino burst even if they are several Mpc away, because the total yield in neutrinos is several orders of magnitude larger than the yield in axions. Second, the $d_2^{(a)}$ value mentioned above is the result of a calculation that considers only axion interactions with hydrogen in water, disregarding the contribution from oxygen. This contribution could potentially impact (i.e., increase) even the order of magnitude of our estimated axion number spectrum. However, its modeling is subject to significantly greater uncertainties than that of the hydrogen component. Third, we only included the processes in \refeq{eq:processes}, whereas additional reactions may further improve prospects for axion detectability. An example are processes of the kind $a N \to Y X$, with $X$ a \che parton and $Y$ a baryon other than a nucleon. We will comment again on all these points in the Outlook section.

With these qualifications in mind, it is clear that the higher the detectability threshold, the shorter the waiting time before an SN event is actually detected—generally through its neutrino burst. In the Milky Way, this relationship can be estimated as $\dot{N}(d) = (d/1~\mathrm{kpc})^2 \times 2.0 \times 10^{-5}~\mathrm{yr}^{-1}$~\cite{Tammann:1994ev}. Integrating over the galactic volume, this estimate is consistent with the expectation of 1 to 3 SN events per century producing neutrino bursts throughout the entire Milky Way \cite{Tammann:1994ev,Li:2010kd,Rozwadowska:2020nab}. As already emphasized, however, not all of these events will also produce detectable axion bursts, because $d_2^{(a)}$ is not even as large as the Milky Way radius: only SN explosions that are sufficiently close-by will produce two detectable bursts, from neutrinos and from axions.

We show $\dot{N}(d)$ in the upper $x$-axis of the panels in \reffig{fig:distance_plots}. The figure thus suggests that one may expect to detect an axion burst for only $\lesssim 10^{-2}$ SN events per century. We emphasize that the $\dot{N}(d)$ scaling estimate above is quite crude; for instance, it does not account for intra-galactic density variations, such as the presence of a core and spiral arms, as opposed to a homogeneous disk. Nevertheless, this estimate should serve as a useful reference. It is therefore crucial to improve the modeling of axion-absorption processes in water that produce \che light, in the hope of uncovering order-of-magnitude enhancement factors.

\section{Outlook} \label{sec:outlook}

The core collapse and cooling that drive the transition from a supernova (SN) to a young neutron star produce a burst of neutrinos. Similarly, this process may also generate fluxes of additional feebly interacting, uncharged particles such as axions. These particles could be detected on Earth by exploiting their interactions with nucleons in reverse, leading to the production of \che light. These interactions take the form $\alpha N \to N^{(\prime)} X$, where the incoming particle $\alpha$ is absorbed by a nucleon $N$ in a water-based detector---such as the planned Hyper-Kamiokande (HK) facility—producing a final-state nucleon $N^{(\prime)}$ and a ``\che parton'' $X$ (such as $e^\pm$, $\gamma$, or $\pi^0$), which initiates an e.m. cascade.

Using Chiral Perturbation Theory, extended to include axions, we calculated the energy spectra for the reactions $a + N \to N + \gamma$ and $a + N \to N + \pi^0$. Our analysis assumes the {\em most general} axion-quark interactions, with coupling strengths constrained either by astrophysical transient observations or by specific theoretical scenarios.

We compared the \che-light spectra induced by axion absorption with those from neutrino interactions. The axion-absorption spectra depend on the axion-nucleon couplings $\Cann, \Capp$, which are fixed according to the different cases of interest mentioned above. We corroborated these findings by assessing the expected reconstruction efficiency for $\pi^0$ spectra based on a realistic HK-like detector simulation. We considered different event categories contributing to $\pi^0$ identification and find an intrinsic efficiency around 80\%, which, while not perfect, is highly promising.

In addition, we examined how the number of detected events depends on the axion-matter couplings, identifying the regions of the $\Cann-\Capp$ parameter space that can be probed based on the expected number of $\pi^0$-initiated signal candidates. Furthermore, we explored how the number of axion-absorption events scales with the number of SN candidates.

Our findings lead to several key conclusions of which we single out the following two: {\em (a)} Even for the QCD axion---which has the weakest interaction strength with matter compared to axion-like particles---the \che light spectrum from neutrino-nucleon interactions is modified in a potentially detectable way.
{\em (b)} Detectability is significantly more promising for the $N + \pi^0$ final state, as its spectrum is an order of magnitude higher in intensity and extends to energies twice as large compared to the $N + \gamma$ counterpart.

The second item is particularly interesting because a larger signal is correlated with a detectable SN event occurring at a larger distance. Typically, SNe are detected through their neutrino bursts, with an estimated rate of 2.1 events {\em per century in the whole Milky Way}. However, {\em not all} of these events will produce detectable axion bursts, as the axion-burst detectability threshold is smaller than the Milky Way's radius. Consequently, only nearby SNe are likely to produce detectable bursts of both neutrinos and axions. Given the rarity of such dual detections, it is crucial to identify additional mechanisms that could enhance axion signals.

To this end, improving the modeling of axion absorption in water \che detectors is essential, as it could reveal potential order-of-magnitude enhancements. Several promising avenues merit further exploration: {\em (i)} On the theoretical side, including the effects of neutrons and the oxygen component of the detector medium could significantly amplify the signal; {\em (ii)} Alternative absorption processes, such as $a + N \to Y + X$, where $Y$ is a baryon other than a nucleon and $X$ produces \che light, also warrant further consideration, as do couplings to other fermions;\footnote{A comprehensive reappraisal of the axion-electron coupling was recently conducted in Ref.~\cite{Fiorillo:2025sln} in the context of emission by SNe.} {\em (iii)} On the experimental side, scaling up detection volumes represents a clear path toward improving sensitivities.

Work is ongoing in directions {\em (i)} and {\em (ii)}.

\section*{Acknowledgments}

We especially thank Micaela Oertel for assistance with SN model files and Pasquale D. Serpico for valuable insights, including on the relation between number spectra for \che partons and \che light. The authors acknowledge useful conversations with Pierluca Carenza, Jeremy Dalseno, Francesco Dettori, Maurizio Giannotti, Giuseppe Lucente, and Diego Mart\'inez-Santos. This research has received funding from the French ANR, under contracts ANR-19-CE31-0016 (`GammaRare') and ANR-23-CE31-0018 (`InvISYble'), that we gratefully acknowledge. LV also acknowledges the Italian Ministry of University and Research (MUR) and the European Union's NextGenerationEU program for support under the Young Researchers 2024 SoE Action, research project `SHYNE', ID: SOE\_20240000025.

\appendix

\section{\boldmath Some details on the computation of the cross sections $\sigma_{a\chep}$} \label{app:sigma}

The computation of the spectrum associated to a generic parton \che $\chep$ requires the knowledge of the total cross section $\sigma_{\alpha\chep}$ associated to the 2-to-2 scattering in \refeq{eq:XN_Nlight}. In this appendix we will focus on the case $\alpha = a$, namely on axion-absorption processes of the form
\be
a \, N_i ~\to~ N_f \,\chep~, 
\ee
where $\chep = \gamma, \pi^0$.

In the reference frame of the laboratory the initial nucleon $N_i$ is at rest, namely $E_{N_i} = m_N$ and $\vec{p}_{N_i} = \vec{0}$. Given that in our setup $m_a \ll m_N,m_{\chep}$, we can safely consider the axion as a massless particle. By starting from the standard expression of the cross section
\be
d\sigma = \frac{1}{4 m_NE_a} \frac{d^3\vec{p}_{\chep}}{(2\pi)^3 2E_{\chep}} \frac{d^3\vec{p}_{N_f}}{(2\pi)^3 2E_{N_f}} \vert \mathcal{M} \vert^2 (2 \pi)^4 \delta^{(4)}(p_a+p_{N_i}-p_{N_f}-p_{\chep})
\ee
and by choosing the direction of the 3-momentum of the axion parallel to the $\hat{z}$-axis, one ends up with the rather simple formula
\be
\frac{\partial\sigma(E_{\chep},\,E_a)}{\partial E_{\chep}} = \frac{1}{32 \pi m_N E_a^2}\vert \mathcal{M} \vert^2 (E_{\chep},\,E_a)~.
\ee
Here $\vert \mathcal{M} \vert^2 (E_{\chep},\,E_a)$ is the square amplitude, that we are going to compute separately for the two different cases $\chep = \gamma$ and $\chep = \pi^0$. Let us finally mention that the angle among the 3-momentum of the axion and the 3-momentum of $\chep$ can be expressed as
\be
\cos\theta_{a\chep} = \frac{2 E_a E_{\chep} + 2 m_N (E_{\chep} - E_a) - m_X^2}{2 E_a \vert \vec{p}_{\chep} \vert}~.
\ee

\begin{itemize}
\item \textbf{$\bm{\chep = \gamma}$ case}
\end{itemize}

In order to compute the square amplitude for $\chep = \gamma$, we consider the contributions coming from diagrams (a) and (b) in \reffig{fig:aN_to_gamma} separately and from their interference. Only the proton gives a non-zero contribution. The final result reads
\begin{equation}
\vert \mathcal{M}[a p \to p \gamma] \vert^2 = \frac{16 \vert C_{app} \vert^2 e^2 m_N^2(p_a \cdot p_{N_f} - p_a \cdot p_{N_i})^2}{f_a^2(p_a \cdot p_{N_f})(p_a \cdot p_{N_i})}~,
\end{equation}
where $C_{app} \equiv C^{(D)}_{app}~ D + C^{(F)}_{app}~ F + C^{(S)}_{app}~ S$ (see \refsec{sec:theory_setup}). Since $p_a \cdot p_{N_i} = E_a m_N$ in the laboratory frame, it is worth mentioning that in the elastic limit, namely for $E_{N_f} = m_N$ and  $p_a \cdot p_{N_f} = E_a m_N$, the amplitude vanishes.

\begin{itemize}
\item \textbf{$\bm{\chep = \pi^0}$ case}
\end{itemize}

In order to compute the square amplitude for $\chep = \pi^0$, we consider the contributions coming from diagrams (d) and (e) in \reffig{fig:aN_to_gamma} separately and from their interference. Both proton and neutron can give a non-zero contribution here. We obtain
\begin{equation}
\vert \mathcal{M}[a N \to N \gamma] \vert^2 = \frac{32 \kappa m_N^2 [2 (p_a \cdot p_{N_i}) (p_a \cdot p_{N_f}) (p_{N_i} \cdot p_{N_f}) - m_N^2 (p_a \cdot p_{N_i})^2 - m_N^2 (p_a \cdot p_{N_f})^2 ]}{F_0^2 f_a^2(p_a \cdot p_{N_f})(p_a \cdot p_{N_i})}~,
\end{equation}
where $\kappa \equiv  \vert C_{aNN} \vert^2 (D+F)^2 / 4$ and $C_{aNN} \equiv C^{(D)}_{aNN} ~ D + C^{(F)}_{aNN} ~ F+ C^{(S)}_{aNN} ~ S$ (see Section \ref{sec:theory_setup}). Also this amplitude vanishes in the elastic approximation.

\section{\boldmath Upper bounds on axion couplings with nucleons from SNe and NSs}
\label{app:bounds}

In \reftab{tab:agnCaNN} we collect the upper bounds on the axion couplings with nucleons, $C_{ann}$ and $C_{app}$, inferred by the Raffelt bound calculated within the different SN models we considered. These models subsume alternative choices for thermodynamic parameters, also listed in the table. In turn, NS cooling data \cite{Buschmann:2021juv} imply the constraints are $|C_{app}| < 1.16$ and $|C_{ann}| < 1.22$ for $f_a=10^9~\GeV$. Within each given SN model, we take the strongest between the SN bound from \reftab{tab:agnCaNN} and the constraints from NS cooling.
\begin{table}[t]
\begin{centering}
\renewcommand{\arraystretch}{1.4}
\captionsetup{justification=centering, font=small}
\setlength{\tabcolsep}{8pt}

\begin{tabular}
{>{\centering\arraybackslash}p{1.5cm}
                >{\centering\arraybackslash}p{1.5cm}|
                >{\centering\arraybackslash}p{2.25cm} 
                >{\centering\arraybackslash}p{2.25cm}| 
                >{\centering\arraybackslash}p{2.25cm} 
                >{\centering\arraybackslash}p{2.25cm}}
\toprule
\rowcolor{lightgray!20}
\multicolumn{2}{c}{} & \multicolumn{2}{c|}{\textbf{DD2}} & \multicolumn{2}{c}{\textbf{SFHo}} \\
\midrule
\rowcolor{lightgray!10}
$T$ [MeV] & $n_B$ &  $ \vert C_{app}\vert$ & $ \vert C_{ann} \vert$ & $ \vert C_{app}\vert$ & $ \vert C_{ann} \vert$  \\
\midrule
30 & $n_{\sat}$ & 2.1 & $2.0$ & 1.8 & 1.7 \\
30 & $1.5\,n_{\sat}$ & $2.0$  & $1.9$ & $1.5$ &1.5\\
40 & $n_{\sat}$ & 0.80 & $0.76$ & 0.67 & $0.64$ \\
40 & $1.5\,n_{\sat}$ & 0.76 & $0.74$ & 0.59& $0.56$ \\
\bottomrule
\end{tabular}
\end{centering}
\caption{
Constraints on the absolute values of the axion couplings to nucleons $\vert C_{app}\vert$, $ \vert C_{ann}\vert$ for $f_a=10^9~\GeV$ for different EOS and for different thermodynamic conditions inside the SN core.
}
\label{tab:agnCaNN} 
\end{table}

\section{Reconstructed event breakdown categorized by SN models and axion interaction constraints}\label{app:T_30_40}

In \reftabs{tab:T30}{tab:T40}, we present the counterparts of \reftab{tab:reco_T_30_40}, where the different event categories are further subdivided based on the two SN models considered, as well as the three alternative ways of constraining the relevant axion interactions. These cases---agnostic, KSVZ, and DFSZ---are outlined in the itemization on page \ref{page:models}. The row labeled ``total'' refers to the detector-level values. The respective truth-level totals correspond to the values in \reffig{fig:money}, properly normalized, and are not explicitly reported.

\begin{table}
    \centering
    \renewcommand{\arraystretch}{1.2}
    \setlength{\tabcolsep}{8pt}
    \begin{tabular}{l c c c | c c c | c}
        \toprule
        & \multicolumn{7}{c}{\bm{$T=30~\MeV$}} \\
        \cmidrule(lr){2-8}
        \textbf{Event} \\ \textbf{Category}  
        & \multicolumn{3}{c|}{\textbf{DD2 and} $\bm{a p \rightarrow p\pi^0}$}  
        & \multicolumn{3}{c|}{\textbf{SFHo and} $\bm{ap \rightarrow p\pi^0}$}  
        & $\bm{\overline{\nu}p \rightarrow ne^+}$ \\
        \cmidrule(lr){2-7}
        & \textbf{Agn.} & \textbf{KSVZ} & \textbf{DFSZ} 
        & \textbf{Agn.} & \textbf{KSVZ} & \textbf{DFSZ} &  \\
        \midrule
        {2-ring $\pi^0$-like} & 90.0 & 1.8 & 9.1 & 135.6 & 2.8 & 13.7 & $<10^{-3}$ \\
        {1-ring $\pi^0$-like} & 19.5 & 0.4 & 2.0 & 29.3 & 0.6 & 2.9 & $<10^{-3}$ \\
        {1-ring e-like} & 11.2 & 0.2 & 1.2 & 17.3 & 0.3 & 1.8 & $2\cdot10^{-2}$ \\
        {1-ring $\mu$-like} & 0.9 & $2\cdot10^{-2}$ & 0.1 & 1.5 & 0.1 & 0.2 & $<10^{-3}$ \\
        {2-ring other} & 8.1 & 0.2 & 0.9 & 13.0 & 0.3 & 1.2 & $<10^{-3}$ \\
        \midrule
        \textbf{Total} & 129.7 & 2.6 & 13.3 & 196.7 & 4.1 & 19.7 & $2\cdot10^{-2}$ \\
        \bottomrule
    \end{tabular}
    \caption{Same as \reftab{tab:reco_T_30_40}, but further categorized according to the two SN models considered, \DDmod and \SFmod, as well as the three alternative methods for constraining the relevant axion interactions: agnostic (constrained solely from data), KSVZ, and DFSZ, as discussed on page \ref{page:models}.}
    \label{tab:T30}
\end{table}
\begin{table}[h]
    \centering
    \renewcommand{\arraystretch}{1.2}
    \setlength{\tabcolsep}{8pt}
    \begin{tabular}{l c c c | c c c | c}
        \toprule
        & \multicolumn{7}{c}{\bm{$T=40~\MeV$}} \\
        \cmidrule(lr){2-8}
        \textbf{Event} \\ \textbf{Category}  
        & \multicolumn{3}{c|}{\textbf{DD2 and} $\mathbf{ap \rightarrow p\pi^0}$}  
        & \multicolumn{3}{c|}{\textbf{SFHo and} $\mathbf{ap \rightarrow p\pi^0}$}  
        & $\mathbf{\overline{\nu}p \rightarrow ne^+}$ \\
        \cmidrule(lr){2-7}
        & \textbf{Agn.} & \textbf{KSVZ} & \textbf{DFSZ} 
        & \textbf{Agn.} & \textbf{KSVZ} & \textbf{DFSZ} &  \\
        \midrule
        {2-ring $\pi^0$-like} & 131.9 & 16.5 & 83.4 & 113.0 & 26.2 & 117.3 & $<10^{-3}$ \\
        {1-ring $\pi^0$-like} & 45.5 & 6.2 & 32.2 & 41.2 & 9.7 & 47.4 & $<10^{-3}$ \\
        {1-ring e-like} & 30.5 & 4.5 & 23.9 & 27.8 & 6.8 & 33.8 & $2\cdot10^{-2}$ \\
        {1-ring $\mu$-like} & 3.0 & 0.4 & 2.0 & 2.4 & 0.6 & 2.9 & $<10^{-3}$ \\
        {2-ring other} & 11.9 & 1.5 & 7.7 & 10.3 & 2.3 & 11.5 & $<10^{-3}$ \\
        \midrule
        \textbf{Total} & 222.8 & 29.1 & 149.1 & 194.7 & 44.6 & 212.9 & $2\cdot10^{-2}$ \\
        \bottomrule
    \end{tabular}
    \caption{Same as \reftab{tab:T30}, but for $T = 40~\MeV$.}
    \label{tab:T40}
\end{table}

\bibliographystyle{JHEP}
\bibliography{bibliography}

\end{document}